\documentstyle[12pt,epsf]{article}

\begin{document}

\begin{titlepage}
\begin{flushright}
  KUNS-1582\\[-1mm]
  {\tt hep-ph/9906549}
\end{flushright}

\vspace*{5ex}
\begin{center}
{\large\bf
  Brane fluctuations and suppression of Kaluza-Klein mode couplings}
\vspace{5ex}

Masako {\sc Bando},$^{1,}$\footnote{E-mail: bando@aichi-u.ac.jp}
Taichiro {\sc Kugo},\footnote{E-mail: kugo@gauge.scphys.kyoto-u.ac.jp}
Tatsuya {\sc Noguchi},\footnote{E-mail: 
  noguchi@gauge.scphys.kyoto-u.ac.jp} and 
Koichi {\sc Yoshioka}\footnote{E-mail:
  yoshioka@gauge.scphys.kyoto-u.ac.jp}

\vspace{1ex}

{\it
$^1$Physics Division, Aichi University, Aichi 470-0296, Japan\\
Department of Physics, Kyoto University, Kyoto 606-8502, Japan
}

\end{center}
\vspace*{8ex}

\begin{abstract}
In higher dimensional models where the gauge and gravity fields live
in the bulk and the matter fields only in a brane, we point out the
importance of the brane (transverse) coordinate modes, which are the
Nambu-Goldstone bosons appearing as a result of spontaneous breaking
of the translation symmetry. The brane recoil effect suppresses the
couplings of higher Kaluza-Klein modes to the matter, and gives a
natural resolution to the divergence problem caused by the exchange of
infinitely many Kaluza-Klein modes. 
\end{abstract}
\end{titlepage}
\setcounter{footnote}{0}

Recently, the existence of large extra dimensions beyond the usual
four dimensions has been intensively investigated. This possibility
can considerably lower the fundamental scale (the Planck 
scale \cite{anto} and/or gauge unification scale \cite{DDG}) to the
scales which are accessible in near future experiments due to the large
volume factors and the effects of Kaluza-Klein (KK) excited states. It
is also interesting that this theoretical assumption can be applied
to models which may explain various phenomenological problems, such as 
fermion mass hierarchy \cite{DDG,fermion}, neutrino 
physics \cite{neutrino}, supersymmetry breaking \cite{susy}, cosmology
and astrophysics \cite{astro}, and so on. In these scenarios, our
world lies on a brane like a D-brane, an orientifold plane, etc.,
embedded in larger dimensions \cite{DDG,brane}.

For charged particles living on the brane which couple to a gauge
field in the bulk, there has been a puzzling paradox: the amplitude
for the two charged-particle scattering in the brane, caused by the
gauge boson exchange, {\it diverges} \/ if the dimensions transverse
to the brane are greater than one. Indeed, the gauge interaction is
naively supposed to take the form
\begin{equation}
  \int d^4xd^\delta y\; g\bar\psi(x)\gamma^\mu\psi(x)\delta^\delta(y)
  A_\mu(x,y) \,=\, \int d^4x\; g\bar\psi(x)\gamma^\mu\psi(x)
  A_\mu(x,0),
  \label{ffA}
\end{equation}
where $x$ and $y=(y^1,\cdots,y^\delta)$ denote the coordinates of the
directions parallel and transverse to the brane, respectively, and the
brane is assumed to lie at $y=0$.  If we assume, for simplicity, that
the transverse directions are all compactified into tori with a common
radius $R$, then the gauge field is expanded into the KK modes
labeled by $n$:
\begin{equation}
  A_\mu(x,y) \,=\, \sum_{n} A^{(n)}_\mu(x)\,e^{in\cdot y/R}.
  \label{mode}
\end{equation}
Then, all the KK modes $A_\mu^{(n)}$ couple to the charged 
particle $\psi$ with {\it equal} \/ strength $g$, and the amplitude for
the two particle scattering is given by
\begin{equation}
  g^2 \sum_{n}\frac{1}{-p^2+n^2/R^2}\,.
  \label{amp}
\end{equation}
But this sum over $n$ diverges when the number of the transverse
dimensions $\delta$ is larger than or equal to 2. 

This is very strange. It is merely a tree-level amplitude and the
divergence cannot be renormalized by any means. In the recent analyses
on the experimental implications of these KK 
modes \cite{exp-gra,exp-cos,exp-vec}, the sum has simply been cut off
at string scale $M_s$; $|n/R|\leq M_s$\@. The string scale here means
the energy scale at which the detailed nature of the brane becomes 
relevant and the effective field theory description on the brane
breaks down. This procedure, however, still seems strange. It is true
that there exists such a built-in cutoff $M_s$, or the ultimate energy
scale, in the present effective theory. But it is not generally true
that the above expression (\ref{amp}) for the scattering amplitude
remains valid until such an ultimate energy scale.

We here note that {\it any type of brane necessarily fluctuates} \/ 
whether it originates from string theory or not. This is because there
cannot exist a {\it rigid} \/ body in the relativistic
theory. Therefore there should always exist scalar fields $\phi(x)$
which stand for the coordinates of the brane in the transverse
dimensions, i.e., the point $x$ in the brane occupies the point
$(x,\,y{=}\phi(x))$ in the bulk. These scalar fields are
Nambu-Goldstone (NG) bosons which appear as a result of spontaneous
breaking of the translational symmetry by the presence of the brane. 

In this letter, we argue that if the effect of brane fluctuations,
namely, the effect of these NG bosons, is correctly included, the
contributions of the higher KK modes are automatically suppressed by
an exponential factor and then the above problematic divergence will
not actually arise.

First, we discuss the effective low-energy interactions of this NG
boson field $\phi(x)$. (Hereafter, for simplicity, we consider the
five-dimensional case, i.e., $\delta=1$, but it is straightforward to
include more numbers of extra dimensions.) \ The couplings 
of $\phi(x)$ to the fields living only on the brane are given through
the induced metric (or induced vierbein) on the brane as is well-known
in string theory and described, e.g., in 
Ref.\ \cite{sundrum}\@. Moreover, when some fields living in the bulk
couple to the fields on the brane, $\phi(x)$ also appears in the
argument $(x,\phi(x))$ of the bulk fields. For instance, the above
naive interaction term (\ref{ffA}) should be replaced by
\begin{equation}
  \int d^4x\; g\bar\psi(x)\gamma^\mu\psi(x)\, A_\mu(x,\phi(x)) \,=\,
  \int d^4x \sum_n g\bar\psi(x)\gamma^\mu\psi(x)\, A^{(n)}_\mu(x)\, 
  e^{i\frac{n}{R}\phi(x)}\,,
  \label{int}
\end{equation}
where the KK mode expansion (\ref{mode}) is substituted on the
right-hand side. Precisely speaking, this is an approximate expression
for the interaction, and the exact expression including the induced
vierbein has more complicated dependence on $\phi(x)$. By examining
the exact expression, however, we can confirm that the 
interaction (\ref{int}) is indeed the leading term in neglecting all
the gravitational and derivative couplings 
of $\phi(x)$ \cite{vierbein}\@. The kinetic term of $\phi(x)$ comes
from the determinant of the induced metric, i.e., the Nambu-Goto
action $\int d^4x (-\tau_4)\,\sqrt{-g}$, which takes the form
\begin{eqnarray}
  \int d^4x\,\left(-\tau_4 +\frac{\tau_4}{2}\partial_\mu\phi(x)
    \partial^\mu\phi(x) +\cdots\right) 
  \label{kinetic}
\end{eqnarray}
on the flat background. Here, $\tau_4\equiv f^4/4\pi^2$ is the brane
tension and the inverse $f^{-1}$ gives a characteristic length as to
how much the brane can fluctuate.

The form of the gauge interaction (\ref{int}) might still seem to
imply the equal couplings for all the KK modes $A_\mu^{(n)}$. However, 
it is deceptive. In a perturbation theory framework, it is appropriate
to rewrite the exponential factor $\exp(i{n\over R}\phi(x))$ into the
normal ordered form referring to the free kinetic term  of $\phi$ 
in Eq.\ (\ref{kinetic})\@. Then the interaction term (\ref{int}) takes
the form
\begin{eqnarray}
  \int d^4x \; g\bar\psi(x)\gamma^\mu\psi(x)\, A^{(n)}_\mu(x)\,
  e^{-\frac{1}{2}\left(\frac{n}{R}\right)^2 \Delta(l_s)}
  :e^{i\frac{n}{R}\phi(x)}:\,. 
  \label{normal}
\end{eqnarray}
Here $\Delta$ is the free propagator of $\phi$
\begin{equation}
  \Delta(x-y)\,\equiv\,\left\langle \phi(x)\phi(y)\right\rangle \,=\,
  \frac{1}{f^4}\cdot \frac{1}{-(x-y)^2}\,.
\end{equation}
Since the present effective theory is valid only at scales larger 
than $l_s=M_s^{-1}$, the propagator $\Delta(x)$ with $|x|\leq l_s$ is
understood to be $\Delta(l_s)=1/f^4l_s^2$, and we have replaced the
infinite $\Delta(0)$ by the value $\Delta(l_s)$ 
in Eq.\ (\ref{normal})\@. This form of the interaction term
(\ref{normal}) now implies that the effective coupling $g_n$ of the
level $n$ KK mode to four-dimensional fields is actually suppressed
exponentially:
\begin{eqnarray}
  g_n \,\equiv\, g\cdot 
  e^{-\frac{1}{2}\left(\frac{n}{R}\right)^2 \frac{M_s^2}{f^4}}\,.
\end{eqnarray}
Note that the origin of this suppression is a recoil effect of the
brane. This is easily seen if we note that the effective 
couplings $g_n$ can also be written as
\begin{eqnarray}
  g_n \,=\, g\cdot\langle 0|\, e^{i\frac{n}{R}\phi(x)}\, |0 \rangle
\end{eqnarray}
in the cutoff theory. In this expression $\left|{0}\right\rangle$ is
the (free theory) ground state of $\phi$ field corresponding to a
Gaussian wave functional peaked at $\phi(x)=0$. Remembering 
that $\phi(x)$ is the transverse coordinate, the 
operator $e^{i\frac{n}{R}\phi(x)}$ is just like the vertex operator in
the string theory and gives transverse momentum $n/R$ to the brane
around the point $x$. Hence, $\,e^{i\frac{n}{R}\phi(x)} |0\rangle$
represents the recoiled state of $\phi$ by the absorption (emission)
of the KK mode $A^{(n)}_\mu$ carrying transverse 
momentum $n/R$ ($-n/R$). Thus the 
amplitude $\,\langle0|\, e^{i\frac{n}{R}\phi(x)}\, |0 \rangle\,$ can
be viewed as a probability of containing the original 
state $|0\rangle$ in the recoiled 
state $e^{i\frac{n}{R}\phi(x)} |0\rangle$. As is clear from this view, 
the suppression becomes stronger for higher KK modes since larger
deformation of the brane occurs, and on the other hand, in case of 
the {\it stiff} \/ brane possessing large $f$, the suppression is weak.

Now, with this new form of interaction term (\ref{normal}), let us
reconsider the problem of the charged-particle scattering amplitude
discussed above. With the coupling (\ref{normal}), the contribution
of the level $n$ KK modes is estimated as
\begin{eqnarray}
  {\cal A}_n &=& g_n^2\cdot \langle A^{(n)}_\mu(x) A^{(-n)}_\nu(0)
  \rangle\, \langle\, :e^{i\frac{n}{R}\phi(x)}:\,
  :e^{-i\frac{n}{R}\phi(0)}:\, \rangle \nonumber \\[1mm]
  &=& g^2\, e^{\left(\frac{n}{R}\right)^2
  (\Delta(x)-\Delta(l_s))}\cdot \langle A^{(n)}_\mu(x) A^{(-n)}_\nu(0)
  \rangle\,.
  \label{An}
\end{eqnarray} 
Note that we here have a factor of effective coupling function
$g^2e^{\left(\frac{n}{R}\right)^2(\Delta(x)-\Delta(l_s))}$. It has an
interesting property that it interpolates from the suppressed value to
the ``bare'' value:
\begin{equation}
  g^2\, e^{\left(\frac{n}{R}\right)^2(\Delta(x)-\Delta(l_s))} \,=\, 
  \left\{
    \begin{array}{ll}
      g_n^2 & {\rm for}\; |x|\gg l_s, \\
      g^2   & {\rm for}\; |x|\simeq l_s.
    \end{array} \right.
\end{equation} 
When we measure the effective 4-fermion coupling at energy scale much
smaller than the cutoff scale $M_s$, therefore, the KK modes
contribute to it as if they have the suppressed gauge 
couplings $g_n = g\cdot\exp[-\frac{1}{2}(n/R)^2(M_s^2/f^4)]$\@. With
this exponential factor damping rapidly with $n$, the higher KK mode
contributions become smaller and the infinite sum of those modes
surely converges as it should be.\footnote{A similar exponential
  factor is also discussed in a string framework with orbifold
  compactification \cite{sup}.}
Note that this convergence remains to hold even when the transverse
dimensions are greater than one, since the exponential damping factor
appears for each extra dimension.

When the brane is hard enough, $f\sim M_s$, the cutoff for the KK
mode sum in the present treatment reduces to the usual naive cutoff
at $|n/R|\sim M_s$ because $g_n/g$ remains to be $\sim1$ until the 
highest KK modes with mass $\sim M_s$\@. However, since we now wish to
consider the effects of the brane fluctuation, we 
assume $f < M_s$. For example, if we take $f \sim R^{-1}\,(\ll M_s)$,
then $g_n/g \sim e^{-n^2 (M_sR)^2}$ which is almost zero for all
non-zero KK modes. In this way, owing to this large exponential
suppression factor with moderate value of brane tension $f$, all the
KK contributions from the tree-level graphs can be suppressed.

We discuss some phenomenological implications of this suppression
factor. Suppose that the standard model gauge fields live in the bulk
while the quarks and leptons live on the brane. Then, as we have
calculated above, the effective 4-Fermi coupling constant $G_F$
receives corrections from the KK excitation modes. The correction is
dominated by the first mode contributions ${\cal A}_{\pm 1}$ in 
Eq.\ (\ref{An}) since the higher modes are further suppressed by the
exponential factors, so that the KK mode correction $\Delta G_F$ to
the 4-Fermi coupling constant can be estimated roughly as
\begin{equation}
  \frac{\Delta G_F}{G_F}\,\sim\, 
  \frac{{\cal A}_{+1}+{\cal A}_{-1}}{{\cal A}_0} \,=\, 
  2\frac{g^2_{n{=}1}}{g^2}\frac{M_W^2}{M_W^2+1/R^2}\,.
  \label{corr}
\end{equation}
Since the standard model prediction precisely agrees with experiments,
the KK mode correction (\ref{corr}) must be small, say
${\raise2pt\hbox{$<$}}\kern-9pt\lower4pt\hbox{$\sim$}\,
O(10^{-2})\,$ \cite{Gf}\@. We consider $R^{-1}$ in the range 
$O(100\,{\rm GeV}) < R^{-1} < M_s$, then the constraint  
$\Delta G_F/G_F < 10^{-2}$ reads 
\begin{eqnarray}
  2M_W^2 R^2\cdot e^{-\frac{M_s^2}{R^2f^4}} \,<\, 10^{-2}\,.
  \label{const}
\end{eqnarray}
This gives a weak constraint to the brane tension $f$. We show in
Fig.\ 1 the allowed region of $f$ for several values of
$M_s$. Clearly, since $\exp[-(M_s^2/R^2f^4)] < 1$, no constraint
appears if $R^{-1}$ is larger than $10\sqrt{2}M_W\sim 1.1$ TeV\@. Even 
for $R^{-1}$ less than this, Fig.\ 1 shows that the constraint
(\ref{const}) can be satisfied for any $M_s$ provided that $f$ is
chosen suitably small. Therefore, if the brane fluctuation is taken
into account, the constraints on the extra dimensions ($R, M_s$) so
far obtained can be substantially loosened. It is clear from the above
discussion that when there are $\delta\, (\geq2)$ extra dimensions,
the suppression factor becomes $(g_n/g)^{2\delta}$ and the constraint
of $f$ is further weaker.

This suppression of the KK mode couplings to four-dimensional fields
also works in the loop diagrams as well as in the tree diagrams. This
effect may be serious and forces reconsideration of the scenarios
proposed so far. Consider, for example, the model proposed in 
Ref.\ \cite{DDG} in which the minimal supersymmetric standard model
gauge fields and Higgs doublets live in the bulk and the other matter
fields are confined on our four-dimensional brane. In this model, the
three gauge couplings evolve up with power-law behavior and rapidly
unify at rather low-energy scale. In addition, Yukawa couplings were
also expected to have the power-law running behaviors, which could
give an interesting origin of the fermion mass hierarchies. However,
the power-law running behaviors were due to the contributions of KK
modes of gauge and Higgs fields in the loop diagrams. If we take the
effect of the brane fluctuation into account, the Yukawa 
couplings {\it no longer} \/ show the power-law running,
unfortunately, since the relevant loop diagrams always contain the
couplings between bulk fields and four-dimensional ones which are
suppressed exponentially for the higher KK modes. On the other hand,
since the gauge fields live in the bulk and their self-interactions
are totally irrelevant to the brane fluctuation, the running of the
gauge couplings still obeys the power law, and the TeV-scale grand
unification scenarios can remain valid.

Finally, we comment on the suppression of the contributions from KK
gravitons. The effects of the KK gravitons have been expected to be
observable in the low-scale quantum gravity 
scenarios ($M_s \sim O({\rm TeV})$) and many authors have investigated
the relevant phenomena observable in the near future collider
experiments and in astrophysics and 
cosmology \cite{exp-gra,exp-cos}\@. Similarly to the case of the
bulk gauge fields, the contributions of the higher KK gravitons are
also suppressed. But this time the compactification radius $R$ should
be very large: $O$(mm) -- $O$(fm), depending on the number of the
compactified extra dimensions, $\delta$. Because of this large radius,
the mass spectrum of the KK modes is nearly continuous and the sum
over the KK momenta can be replaced by an integration. For example, 
consider the process of matter scattering on the brane by the exchange
of an infinite number of KK gravitons. The amplitude ${\cal M}$ takes
the following form, omitting the numerator factors unimportant to the
present discussion:
\begin{eqnarray}
  \sum_n \frac{1}{M^2_P}\, \frac{e^{-\left(\frac{n}{R}\right)^2
      \frac{M^2_s}{f^4}}}{\left(\frac{n}{R}\right)^2-t}
  &=& \frac{R^\delta}{M_P^2}\int d^\delta k\,e^{-k^2\frac{M^2_s}{f^4}} 
  \int_0^\infty d\tau\,e^{-(k^2-t)\tau} \nonumber\\
  &=& \frac{\pi^\frac{\delta}{2}}{M_s^{\delta+2}}\int_0^\infty d\tau\, 
  e^{t\tau} \Bigl(\tau+\frac{M^2_s}{f^4}\Bigr)^{-\frac{\delta}{2}},
  \label{graamp}
\end{eqnarray}
where $t\;(<0)$ is the momentum transfer ($|t| \gg R^{-2}$), $1/M_P^2$
is the graviton coupling constant, and the 
relation $M_P^2=M_s^{\delta+2}R^\delta$ has been used. Note that the
presence of the exponential suppression 
factor $\exp[-(n/R)^2(M_s^2/f^4)]\,$ has made it possible to integrate
over the KK momentum without artificial 
cutoff. When $\sqrt{|t|}\ll f^2/M_s$, the amplitude (\ref{graamp})
approximately becomes
\begin{eqnarray}
  {\cal M} &\sim& \frac{1}{M_s^4} \ln \left(\frac{f^4}{-tM_s^2}\right)
  \qquad (\delta=2), \\
  &\sim& \frac{1}{M_s^{\delta+2}}
  \left(\frac{f^2}{M_s}\right)^{\delta-2} \qquad (\delta> 2)\,.
\end{eqnarray}
In the case of $f\simeq M_s$, this expression reduces to the usual
result by the simple cutoff taking no account of the brane
fluctuation. If the tension $f$ is smaller than $M_s$, however, we see
that the cross section $\sigma$ for this process is suppressed as 
\begin{eqnarray}
  \sigma \,\simeq\, \left(\frac{f}{M_s}\right)^{4(\delta-2)}\!\!\times 
  \sigma(f=M_s)
\end{eqnarray}
for $\delta>2$\@. Because of the large power $4(\delta-2)$, the
anticipated observable effects from KK gravitons are, unfortunately,
strongly suppressed with moderately small values of $f/M_s$, and then
no new constraint on the fundamental scale $M_s$ may be imposed from
these arguments. The case of two extra dimensions, $\delta=2$, is
exceptional, where the suppression effect by the brane fluctuation is
only logarithmic and the effects of KK gravitons may still be
observable.

In conclusion, we have discussed the effects of brane reactions to
the emission and absorption of the KK modes. With a generic value of
the brane tension $f<M_s$, the couplings of higher KK modes to the
matter on the brane are exponentially suppressed by the brane recoil
effect and then the problematic divergence in summing up those
contributions is automatically avoided. When the brane tension $f$
becomes smaller, the KK mode couplings becomes weaker and hence the
detection of their signatures becomes more difficult. This might seem
unwelcome from a phenomenological viewpoint. However, since $1/f$
plays the role of the coupling constant of the brane coordinate NG
mode $\phi$, it will become easier to detect $\phi$ produced copiously
in various processes \cite{vierbein}\@. We hope the future experiment
will check the existence of $\phi$, which is a direct evidence for our
world to lie on a brane.

\subsection*{Acknowledgments}

The authors would like to thank K.\ Hashimoto, T.\ Hirayama, 
H.\ Kawai and M.\ Yamaguchi for many valuable discussions and
comments. M.~B.\ and T.~K.\ are supported in part by the Grants-in-Aid
for Scientific Research No.\ 09640375 and No.\ 10640261, respectively,
from the Ministry of Education, Science, Sports and Culture, Japan,
and K.\ Y.\ is supported in part by the Grant-in-Aid for JSPS Research
Fellowship.

\newpage

\begin{figure}[htbp]
  \begin{center}
    \leavevmode
    \epsfxsize=10cm \ \epsfbox{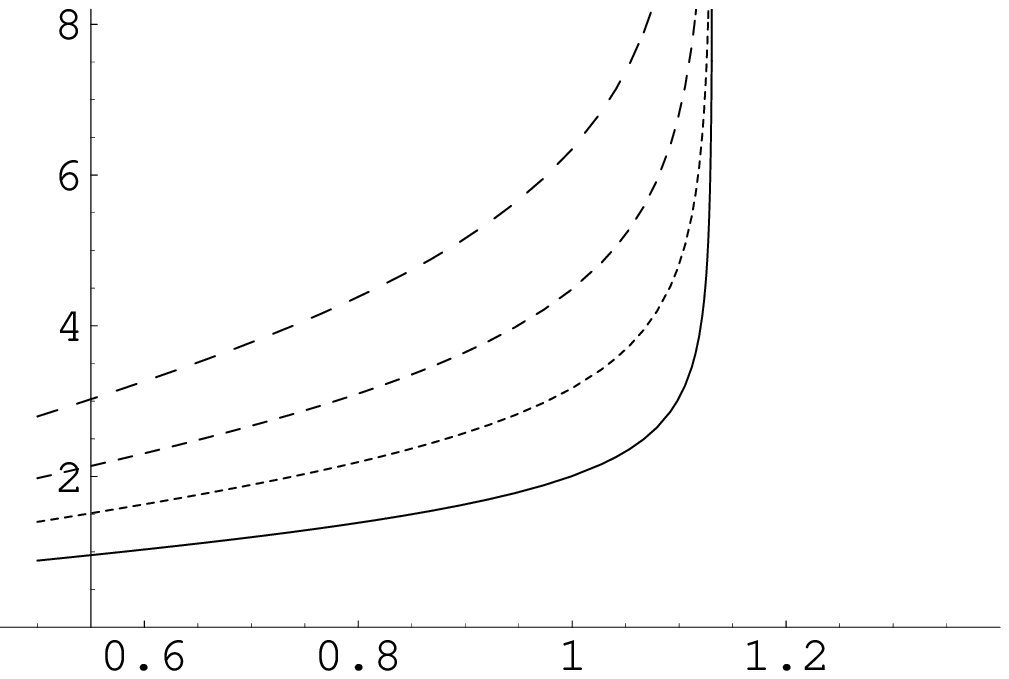}
    \put(-270,212){$f$ [TeV]}
    \put(10,20){$R^{-1}$ [TeV]}
    \put(-85,45){\fbox{Allowed}}
    \put(-220,140){$M_s=20$ TeV}
    \put(-156,108){10}
    \put(-132,92){5}
    \put(-112,77){2}
    \caption{Typical allowed region for brane tension $f$ from the
    measurement of the Fermi constant. The region below each line is
    allowed for each value of $M_s$\@. For $R^{-1}\; 
    {\raise2pt\hbox{$>$}}\kern-9pt\lower4pt\hbox{$\sim$}\;\, 1.1$ TeV,
    there exists no constraint on $f$\@.}
  \end{center}
\end{figure}

\end{document}